\documentclass[10pt, conference]{IEEEtran}
\def\IEEEbibitemsep{0pt plus .5pt}
\usepackage[font={footnotesize}]{caption}
\usepackage[ruled]{algorithm2e} 
\usepackage{listings}
\usepackage{amsmath}
\usepackage{graphicx}
\usepackage{tabularx,ragged2e,booktabs,caption}
\usepackage{verbatim}
\usepackage[T1]{fontenc}
\usepackage{makecell}
\usepackage[usenames, dvipsnames]{color}
\usepackage{varioref}
\usepackage{multirow}
\usepackage{hhline}
\DeclareCaptionFormat{listing}{{\parbox{\dimexpr\linewidth-1\fboxsep\relax}{#1#2#3}}}

\ifCLASSINFOpdf
\else
\fi
\begin{document}
%
\title{Micro-level Modularity of Computaion-intensive Programs in
Big Data Platforms: A Case Study with Image Data}
\vspace{-.10cm}

\author{\IEEEauthorblockN{ Amit Kumar Mondal ~ Banani Roy ~ Chanchal K. Roy ~ Kevin A. Schneider}
	\IEEEauthorblockA{University of Saskatchewan, Canada\\
		\{amit.mondal, banani.roy, chanchal.roy, kevin.schneider\}@usask.ca}
}
\vspace{-.10cm}


%


\maketitle

\begin{abstract}

With the rapid advancement of Big Data platforms such as Hadoop, Spark, and Dataflow, many tools are being developed that are intended to provide end users with an interactive environment for large-scale data analysis (e.g., IQmulus). However, there are challenges using these platforms. For example, developers find it difficult to use these platforms when developing interactive and reusable data analytic tools. One approach to better support interactivity and reusability is the use of micro-level modularisation for computation-intensive tasks, which splits data operations into independent, composable modules. However, modularizing data and computation-intensive tasks into independent components differs from traditional programming, e.g., when accessing large scale data, controlling data-flow among components, and structuring computation logic. In this paper, we present a case study on modularizing real world computation-intensive tasks that investigates the impact of modularization on processing large scale image data. To that end, we synthesize image data-processing patterns and propose a unified modular model for the effective implementation of computation-intensive tasks on data-parallel frameworks considering reproducibility, reusability, and customization. We present various insights of using the modularity model based on our experimental results from running image processing tasks on Spark and Hadoop clusters.

\end{abstract}


%
\IEEEpeerreviewmaketitle

\section{Introduction}

With the rapid advancement of Big Data platforms, software systems \cite{genap, pairs, tomogram} are being developed that provide end-users' an interactive environment for large-scale data analysis in the area of scientific research, business, governments, and journalism. An interactive environment provides a drag-and-drop facility for composing reproducible computation (workflows/pipelines) from a collection of sub-tasks without much technical knowledge. Big Data platforms such as Hadoop \cite{hadop}, Spark \cite{Spark}, Google Dataflow \cite{data_model}, and so on provide  high-level abstract interfaces for implementing distributed-cluster processing of data using commodity hardware. Recently, a number of researchers \cite{cloud_archi, graywulf, module_soft,iabdt, hbaselarge} focused on developing architectures and frameworks for large-scale data analysis tools utilizing these platforms. Most of these architectural frameworks are adopting workflows and pipelines \cite{traverna, kepler, illumina} for reproducible data analysis tasks according to the user requirements. However, data-storage models, data-structures, data-operations, accessing and visualization of large data are complex to handle. Existing literature \cite{model_deployment} suggest that significant effort is spent developing data processing pipelines. Besides, a recent empirical study \cite{notall} reports that data engineers are facing great difficulties to work with Big Data platforms. In order to reduce the development efforts and providing better programming flexibility, a few studies attempted to develop more abstract and unified programming interfaces (especially in Bioinformatics and GIS research) \cite{sparkseq, biosspark, spatialhadoop, stormcv, kira, hipi} as a layer on top of these platforms (e.g., Hadoop and Spark). However, most of them are still in the development phase (e.g., SparkSeq \cite{sparkseq}), and some of them only implemented and tested a few specific tasks within a certain domain. 
Although, a few works implemented large scale image processing tasks \cite{stormcv, kira, hipi, iabdt}, they did not provide an analysis study about the underlying challenges and solutions of using these platforms for real world image processing pipelines. Moreover, common unified frameworks are not readily available to implement reproducible image processing pipelines covering a wide area with Big Data platforms. A few researchers \cite{tomogram, highregistration} have attempted to tune cluster resources for performance optimization of the tasks. Nevertheless, resource enhancement may not provide a feasible solution even with the availability of enough computing power. Therefore, interactive large-scale data-analysis with a Big-Data platform is still a challenging task for the programmers and developers.

Modularization is an important paradigm in software design which provides special program constructs, such as shared data structures or abstract and unified frameworks. Modularization is the action of \emph{"de-composing a system into modules"} \cite{leveragedr}. Moreover, modularization is essential for scalable and interactive application development with Big Data platforms \cite{mrrunner, model_deployment, module_soft}. Our focus is on modularising interactive, data-intensive programs so that they operate effectively on map-reduce frameworks in order to support reusability, reproducibility, and customization. Splitting tasks considering large scale data processing and computation logic reusability may have adverse effects when run on Big Data platforms. However, micro-level modularity has been shown to work successfully on map-reduce frameworks for a number of applications, including machine learning \cite{knnis, bigkmean} and graph data processing \cite{biggraph, graphx}. We are also motivated to support developers of Big Data analytic tools. By separating tasks into further independent micro-components based on data-processing patterns, we hope to develop a unified programming interface that will provide the flexibility for accelerating the development of interactive, re-usable Big Data analytics tools.


Although Big Data platforms hide the complexity of distributed computing, they provide a limited number of methods (e.g., \emph{map, filter, reduce}) for data parallel operations. Adding an extra data processing step with those methods could increase the computation and memory overhead in a significant way. For example, Smith and Albarghouthi \cite{mapreduce} discuss the challenge of partitioning computation with data-parallel operators (\emph{map, filter, reduce}). Due to the complexity of partitioning, they avoid optimization of their technique. In order to reduce working efforts, a few authors \cite{mrrunner, model_deployment} focused on developing frameworks for running and re-deploying modular jobs and a statistical model provided by the users. Unfortunately, none of them conducted extensive study on any effective techniques for modularity to examine the impact of modularity on computation-intensive tasks. Moreover, the mechanism of controlling data-flow among intermediate steps is really important for reproducible computation of large scale data. All things considered, we propose a modularity model and observe the behaviors of different applications in terms of modularization. 
Overall, in our work, we mainly focus on two research questions:

\textbf{RQ1.} How to modularize data and computation-intensive programs to provide a unified abstract framework for developing interactive tools?

\textbf{RQ2.} How does splitting up of run time job-data and processing logic affect the performance of computation-intensive tasks in map-reduce platforms?

In order to answer \textbf{RQ1}, we analyzed various open source image processing tools and state-of-the-art image processing techniques that cover a wide range of tasks. We look into the programming models and data-types that are produced during a full image processing task (some of them are presented in Table \ref{tab:steps}). Then, we categorized the image operations and defined a data processing pattern that is fruitful for modularizing the tasks with Big Data platform. After that, we proposed a micro-level modularity model consisting of four major data-parallel modules each having three core layers (the second layer controls parallel data-flow).
For answering \textbf{RQ2}, we implemented six image processing applications following the proposed modular model. Then we experimented with both the minimal and modularised version with various datasets in a Spark cluster. From the experiment, we found that the task modularization affects system's performance and flexibility of pipeline development. Performance varies case by case with some tasks improving, some decreasing, and others unaffected. For all the cases, it opens up the facility of flexible implementation with data-parallel components. Notably, we also identified the challenges of image processing with data-parallel frameworks from our experimentation. In summary, our case study provides a modularization technique and helpful knowledge-base for interactive tools developers for large scale image processing. Our defined data-pattern and modularity model can be used as a design pattern and design rule in this domain.

The rest of the paper is organized as follows. Section II describes the process of extracting data-processing patterns. Section III presents our proposed modular model. Section IV provides our experimental results. Section V provides discussion and some useful insights from the lesson learned. Section VI describes related work. Finally, section VII presents the conclusion and future work of the paper.

\section{Modularising Data-intensive Tasks}
In our study we focused on image data since a framework that supports various image processing pipelines is not readily available. On the contrary, few abstract frameworks \cite{kira, hipi, icp} are being developed for a few specific image processing applications and most of the Big Data frameworks support workflows for text data processing \cite{spatialhadoop, biosspark}. We conduct our case study following three major strategies: (i) Background and Contextual Analysis, (ii) Data-processing Pattern Extraction, and (iii) Transformation to Data-parallel Components. 

\subsection{Background and Contextual Analysis}
To develop a unified framework, understanding the context is essential. 
To that end, literature review and analysis of various architectures \cite{cloud_archi, graywulf, module_soft,iabdt}, frameworks, tools, techniques, and open-source APIs in the scientific data analysis are essential to determine the exact support needed for the data scientist. Analyzing the recent development strategy of analytic tools for large scale data, we notice that some of the developed applications follow a workflow based modularity architecture \cite{module_soft, graywulf}, whereas others follow a layered architecture \cite{tomogram,iabdt, pairs}. In the workflow based modularity architecture, applications are designed using a special data model which is much different than the traditional model view controller model. For example, the architecture of IQmulus \cite{module_soft}, a GIS data processing system, is heavily dependent on data-analysis workflows. High-level components, job manager, processing services etc. are designed focusing on the on-the-fly workflow compositions. Still users need to learn a considerable amount of script for composing workflows for GIS. Similarly, GrayWulf \cite{graywulf} handles two types of workflows: (i) one is for data manager, and (ii) another is for end-users. The architectural model is based on these workflows composition. However, using GrayWulf, a smaller amount of processed result can be shared and retrieved in the cloud. Another application for image analysis, IABDT \cite{iabdt} followed multi-layer architecture and primarily
used HadoopImageBundle (HIB) for performing basic operations on image data. In a recent study, Roy et al. \cite{cloud_archi} focus on data-centric component development for an application that supports large scale data analysis. Besides, most of the unified frameworks to support applications development as mentioned above followed
specialized data-models for large scale data processing with
distributed clusters. Among existing popular frameworks in the scientific
analysis, SparkSeq is based on Hadoop-BAM \cite{hadoopbam} data frameworks. Hadoop-BAM is created to
solve the issue of map-reduce implementation and attempted to include all data formats in bioinformatics. A unified framework for large scale Geospatial data analysis, SpatialHadoop \cite{spatialhadoop} added three more layers on top of Hadoop to drive efficient
map-reduce based processing of GIS data. KIRA \cite{kira} is
written using SEP library and FITS data model for analyzing the astronomical object. All of the evidence prompt that tool development in Big Data platforms requires different design rule and modularity models. Yet, the common obvious advantages of modularization \cite{leveragedr} in software development are: (i) Easier to Debug and Problem Detection, (ii) Reusable Code, (iii) Readability, and (iv) Reliability. Debugging time is lengthy during the development of Big Data analytic tools. Most of the time, ultimate problems cannot be detected until the application is run on the live cluster with the full set of data. In summary, for large-scale data analysis the following trends are emerging:

(i) Suitable architectural model \cite{cloud_archi, module_soft, iabdt}, 

(ii) Work-flow processing and management \cite{traverna, kepler, module_soft},

(iii) Data-pipelines \cite{illumina}, 

(iv) Data-flow management \cite{data_model, pig, udf_dataflow},

(v) Data-centric decoupling of programs \cite{graphx, biggraph}, 

(vi) Efficient data-storage model \cite{spatialhadoop, pairs}, and 

(vii) Intelligent modularization \cite{mrrunner, model_deployment}. 

\begin{figure}
	\centering
	\vspace{-.3cm}
	\includegraphics[height=1.10in, width=3.2in]{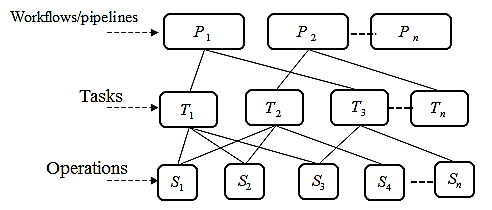}
	\caption{Reproducible workflow composition technique (tasks and operations are reused, algorithms are customized).}
	\vspace{-.4cm}
	\label{workflow_comp}
\end{figure}

The central objective of all of these paradigms is to make scientific computation reproducible \cite{reproducable_computation} with minimal technical knowledge. Figure \ref{workflow_comp} demonstrates how reproducible workflows/pipelines are constructed.  
However, large scale image processing domain requires more focus on all of the above-mentioned directions. In order to understand and develop a knowledge-base, we look into the properties of various open source image processing tools \cite{imagej,htpheno,plantcv,imageharvest, bisque, kira,hipi,stormcv}.

Analyzing source code of the open-source tools facilitate us more intuitive insight about the implementation of real world image analysis applications, programming models, I/O operations, and data entities that analysts, researchers, or end-users might re-use later. We also observe that in image processing tools following attributes are influential (some of them are identified by Heit et al. \cite{model_deployment} in data mining as well): (i) Image pipeline composition, reuse and management, (ii) Image processing workflow modeling language, (iii) Image storage service, (iv) Collaboration between data scientists, (v) Deployment and third party service communication, (vi) Scalability, and (vii) Plugin development and integration.
Another key thing to remember is that data structure and computation model of images are complex and diverse \cite{imagej}. Images ($I$) consist of different data units (8-bit, 16-bit and 32-bit), formats (TIFF, GIF, JPEG, BMP, DICOM, FITS), or dimensions/channels (2D, 3D, 3-channels and so on). Additionally, we found that many algorithms are operated on an individual image except for machine learning/statistical model and template generations. For developing a desired image processing task, programmers and researchers need to experiment with various combinations of techniques and algorithms (hundreds of algorithms are available) along with parameters tuning for each of the canonical operations. Moreover, these operations are tested on large collections of images multiple times and the experimental setup needs to be stored for the future run.
Many core image operations can be found within a popular open-source image processing API called OpenCV \cite{opencv}. Therefore, if a framework can be devised that also facilitates an automatic transformation of iterative operations for a single image into parallel (processing with multiple computers) one for multiple images will be valuable for the data scientists. All of this knowledge-base is useful for unified framework development and reshaping the modularization for Big Data frameworks. In the following sections, we will present analysis study of various image processing tasks to extract data-processing patterns and transforming the concept into the data-parallel framework.

\subsection{Program Synthesis and Extracting Data Processing Patterns} \label{data_pattern}
From the previous discussion it is persuasive that in large-scale data processing, most of the techniques, algorithms, frameworks, and software models are extensively data-centric \cite{cloud_archi, bigdataopportunity, udf_dataflow}. In data-centric development, at first a core-feature model is developed, then data processing patterns are extracted from feature model, targeted technologies and real-world experience, and finally, components are designed and modularized based on the pattern. Therefore,
understanding data-processing patterns is an important part of implementing modularised, split and decoupled data processing applications. 

Our selected algorithms and techniques cover various image processing tasks in plant science, agriculture, biomedical, astronomy, and general computer vision. A total number of 30 applications we analyzed are presented in Table \ref{tab:itasks}. Some of the selected image processing tasks with the major steps and their corresponding produced entities are presented in Table \ref{tab:steps}. 
The high-level steps are shown in Table \ref{tab:steps} of extracting texts \cite{banglaocr} from a video are: gray-scale conversion and noise removal, feature calculation, detecting text areas, then extract texts from the segmented areas of the video images. We notice that produced output of various steps has different data structures in most of the cases. However, it is necessary to figure out an optimal and unified model that might be fitted for a wide area. Most of the image analysis tasks can be divided into four re-usable tasks (please note that here we consider single image analysis task, while two or more image analysis tasks are used to compose a complex pipeline like HtPheno \cite{htpheno}). Many tasks have more than four steps. However, in terms of data and program reusability, for image registration (presented in Table \ref{tab:steps}), matching points calculation ($S_{3}$) and Homography generation ($S_{4}$) can be considered as one logical and independent step. Similarly, in Tomograms generation, refined class objects are reused later, thus $S_{3}$ and $S_{4}$ can be combined into a single step logically.

\begin{table}[h!]
	\begin{center}
		\vspace{-.3cm}
		\captionof{table}{Number of image processing tasks we analyzed in different areas} \label{tab:itasks} 
		\begin{tabular}{ c|c } 
			\hline
			\textbf{Area of Application}
			&
			\textbf{Number of Image processing Applications}
			\\ 	\hline
			Plant and Agriculture & 14 \\ \hline
			Medical and Biology & 5 \\ \hline
			Astronomy & 4 \\ \hline
			General computer vision
			& 7 \\ \hline
	
		\end{tabular}
			\vspace{-.3cm}
	\end{center}
\end{table}

\begin{table}
\scriptsize
	\begin{center}
				\vspace{-.3cm}
		\captionof{table}{Applied techniques and output for each step of various image processing tasks } \label{tab:steps}
		\begin{tabular}{p{.20in}|p{0.60in}|p{0.60in} |p{0.55in}|p{0.60in} } \hline
			
			\textbf{\makecell{ Steps}} & \textbf{Img Classification \cite{img_class}} & \textbf{Img Registration \cite{img_regi}} & \textbf{Text Extraction\cite{banglaocr}} & \textbf{Pattern in Tomograms\cite{tomogram}} \\
			\hline
			\makecell{$S_{1}$} & \makecell{Grey conversion \\ $I_{P}$} & \makecell{Grey conversion \\  $I_{P}$} & \makecell{Decomposition \\  $I_{D}$} & \makecell{Gaussian Trans. \\  $I_{G}$} 
			\\\hline
			
			\makecell{$S_{2}$} & \makecell{Feature extract \\  $V_{F}$=\{$F_{1},..Fn$\}}& \makecell{Metrics Calc. \\ $V_{M}$=\{$M_{1},.M_{n}$\}} & \makecell{Feature extract \\  $V_{F}$} & \makecell{Feature extract \\ $V_{F}$} 
			\\\hline
			
			\makecell{$S_{3}$} & \makecell{Train \& Model \\ $M_{M}$={K-mean}} & \makecell{Matching points\\ $P_{M}$} & \makecell{Train\& Model \\  $M$={Mlp, An}} & \makecell{Cluster objects \\  $OB_{n}$}
			\\\hline
			
			\makecell{$S_{4}$} & \makecell{Grouping \\$C_{n}$ \{n=1,2.. \}}& \makecell{Homography \\ $M_{H}$=[..][..]} & \makecell{Area segments \\ $I_{S}$=\{$S_{1},..S_{n}$\}} & \makecell{Refine Class-objs \\  $C_{n}$} \\\hline
			
			\makecell{ $S_{5}$} & ~~~-- & \makecell{Warping\& align \\ $I_{r}$} &  \makecell{Text extraction \\ $T_{r}$} & \makecell{Generate Tomog. \\  $I_{T}$} 
			\\\hline
			
			\multicolumn{5}{c}{ Here, $I_{p}$ to $I_{g}$ - processed image, $V_{F}$ - feature vectors, $I_{r}$ - result} \\
			
			\multicolumn{5}{c}{$C_{n}$ - list of classes} \\\hline
		\end{tabular}
			\vspace{-.3cm}
	\end{center}
\end{table}
Therefore, analyzing the above-mentioned tools and techniques, we categorize the canonical operations of image analysis tasks into major four steps:

\textbf{Preprocess/conversion ($S_{1}$):} This step is the first and very common for every image analysis pipeline.  This step may produce different kinds of output (such as grayscale image and Canny edge image) based on applied techniques or algorithms, such as Gaussian blurring, wavelet transform, image contrasting, enhanced, noise reduction and so on \cite{opencv}. 

\textbf{Estimate/Extraction ($S_{2}$):} In this step, different kinds of algorithms such as SURF, SIFT, ORB, HOG \cite{opencv} are applied for calculating features, metrics, and key points. However, other texture generation techniques are also employed after the feature and keypoint extraction step. This step produces an array, vector or list type data-structures.

\textbf{Model/Fitting ($S_{3}$):} This step uses extracted features, metrics or composed data for fitting, training or developing models for generating templates based on which final analysis and processing are done.

\textbf{Analysis/Postprocess ($S_{4}$):} This final step mainly produces processed images and analysis result based on the generated template or the model in the model-fitting step. The produced results of this step include matched images, extracted objects, clusters of images, and registered images along with statistical results.   

Such a categorization of the operations based on produced data and computation logic would help developers and programmers to wrap image processing tasks into a common data model and abstract frameworks. Furthermore, program synthesis of the above-mentioned image processing tools, their operations, and I/O operations allowed us to come to a conclusion that produced data in various steps as discussed in Table \ref{tab:steps} can also be saved for later reuse. Consequently, these tasks should be modularised not only for program reuse but also for data entities re-use. 
In summary, the data processing pattern for image processing tasks can be described as follows (as shown in Table \ref{tab:steps}):
\begin{itemize} 
	
	\item Input of first step ($S_{1}$) is generally $\{I, R_{S1}\}$, where I is raw images and $R_{S1}$ is parameters, produced data are processed images $I_{P}$. Parameters may be numeric values, meta-data, vectors, or even raw images.
    
    \item Other two steps ($S_{2}$,$S_{3}$) input are $\{I_{P}, R_{S(2/3)}\}$, where $I_{P}$ is the produced entity of the previous step and $R_{S(2/3)}$ represents parameters of these steps.
    
    \item However, in some cases the input of the last step ($S_{4}$) is  $\{I, I_{P}, D_{M}, R_{S4}\}$; where $D_{M}$ is the model or template generated in the third steps, and $I_{P}$ is the outcome from the first step.
    \item We observe that many cases, $I$ and $I_{P}$ are required to flow and retain up to the last step which is handled with disk storage in localized processing.
    \item For a few cases, images are required to group or bundle during $S_{3}$ and $S_{4}$ (e.g., image registration and panoramic view generation). Likewise, produced results in image processing tasks have various types. Those are a single image, a collection of images or image objects ($I_{R}$), list of string or numeric values ($L_{s}$), collection of matrix or vectors ($L_{M}$), dictionary ($D_{S}$), tuple of lists ($T_{L}$), and so on. 
\end{itemize}

This common pattern is the basis for interactive image analytic tools development for both usual and large scale data. In the next subsequent sections, we will discuss in details how to implement this data-processing patterns into map-reduce frameworks considering a unified programming interface.

\section{Transformation to Data-parallel Modules}
In this section, we discuss how to implement the image processing tasks into modularised and abstract steps in Big Data frameworks following the extracted data-processing pattern. We focus on Apache Spark (with HDFS) implementation which is optimized and the mostly used \cite{mapreduce, knnis} framework. Here, the data-processing pattern serves as modularity properties. We will use many terms and symbols to avoid frequent use of the phrases in our description (many of them are introduced by Smith and Albarghouthi \cite{mapreduce}). 

\subsubsection{Challenges}
Recent works \cite{mapreduce, knnis, graphx,bigkmean, biggraph} with map-reduce frameworks provide firm evidence that map-reduce based implementation is non-trivial for flexible and scalable data processing. Moreover, many applications are yet to a good fit for Big Data platforms using traditional map-reduce techniques due to network induced non-determinism, data shuffling \cite{mapreduce}, and run-time data increment \cite{knnis}. For example, researchers are still working to make KNN more feasible for large data with Big Data platforms \cite{knnis}. Storage files of text and Genome data could be partitioned into further smaller blocks for efficient distributed processing. But data file of each image and associated meta-data is required to treat as a single unit for image processing. Few images among thousands of collection might be corrupted and disrupt the whole processing task. This scenario is also required to handle during large scale processing. However, all the operations in map-reduce based platform (i.e., Spark) should be done with the data parallel components ($\sum _{DP}$): \emph{map(), reduce(), filter(), join(), repartition(), subtractbykey(), count(), collect()} along with \emph{$\lambda-$expressions} (PABS) \cite{mapreduce}. All the image operations cannot be easily paralleled with this platform. When data size is big enough, a single additional operation with $\sum _{DP}$ takes a significant amount of time. Moreover, broadcasting data entity frequently to the worker processes might add further overhead. Consequently, programmers are required to be more careful and thoroughly test with a full dataset.
For reusable computation, each step should be independent in terms of execution, data sharing, and data storing. Having said that, steps should not be divided arbitrarily like usual programming. As we discussed in Section \ref{data_pattern}, raw-data, processed data, and external parameters need to flow from one step to another, and this might increase both memory and time overhead (with the number of steps). Apart from these, handling of various types of produced results (as described in the data processing pattern) requires a well-defined rule to store in a distributed environment.

\subsubsection{Proposed Modularity Model}
\begin{figure}
	\centering
	\includegraphics[height=1.70in, width=3.60in]{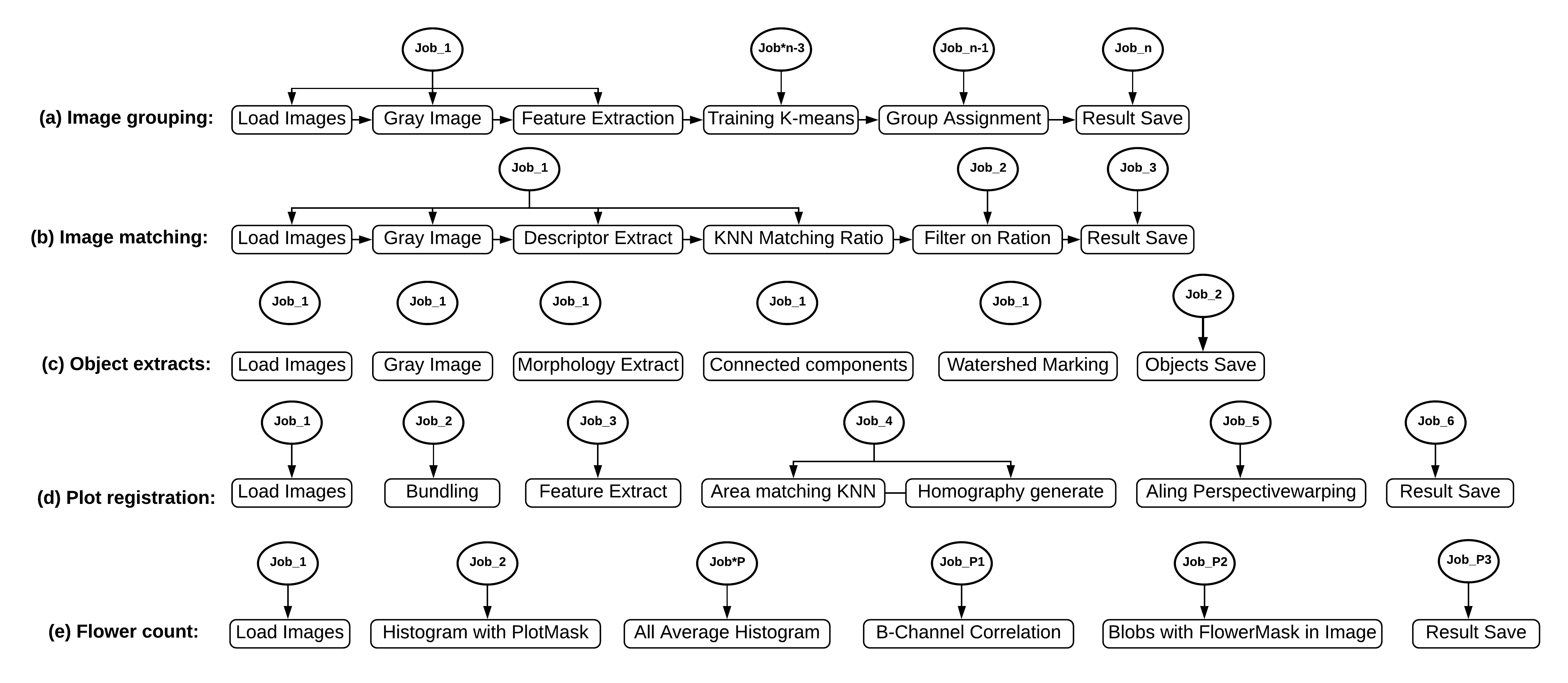}
	\vspace{-.3cm}
	\caption{Modularity options of image processing tasks in the data-parallel framework (more modules mean more data-flow).}
	\vspace{-.4cm}
	\label{modular}
	
\end{figure}

\begin{figure}
	\centering
	\includegraphics[height=2.50in, width=2.80in]{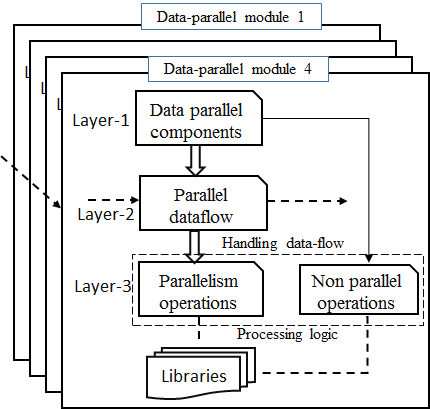}
	\caption{Micro-structure of a data-parallel module.}
	\vspace{-.4cm}
	\label{mdp}
	
\end{figure}

\begin{figure}[t!]
	\centering
	\includegraphics[height=1.75in, width=3.5in]{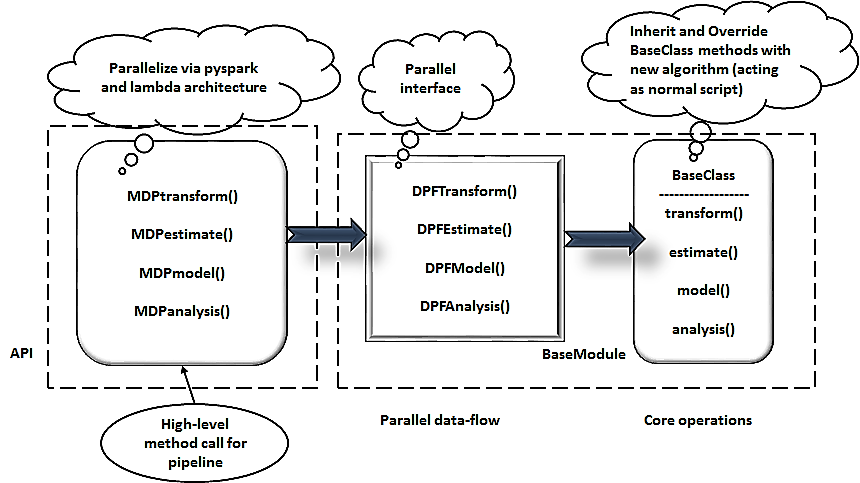}
	\vspace{-.3cm}
	\caption{An example unified interface of image pipelines with Spark.}
		\vspace{-.4cm}
	\label{interface}

\end{figure}

Image processing tasks can be implemented in a various number of modularized steps (one or more)  with data-parallel frameworks as shown in Figure \ref{modular}. Here we introduce data-parallel module, $M_{DP}=\bigcup\limits_{i=1}^{n} DP_{i}$ as a combination of one or more data-parallel components in $\sum _{DP}$. The split into $M_{DP}$ is followed by the corresponding data-processing patterns presented in Section \ref{data_pattern}. From the analysis of data processing patterns of image processing applications, we identified four canonical steps: $S_{1}, S_{2}, S_{3},S_{4}$. A step is a combination of many operations (some of them are canonical also), and it is essential to detect which operations require parallelism and which parts do not. We can represent $S_{i}=\{\sum_{OS}, NP_{S}\}$, where $\sum_{OS}$ represents operations that require parallelism, and $NP_{S}$ represents not parallel. All $\sum_{OS}$ within a $M_{DP}(S_{i})$ should be combined in such a way that the number of $\sum _{DP}$ are minimal (i.e., this rule restricts the modularity of usual computation). However, for a few steps in some cases, run-time data should be partitioned (based on heuristics \cite{knnis}) for further optimization (as shown in Listings \ref{goodflowercount}). A module, $M_{DP}$ must produce a meaningful outcome that can be reused in future either by one of the independent operations in $S_{1}$ to $S_{4}$ or another task (or pipelines). However, a complete Image analysis task could be implemented with one or two minimal steps in map-reduce frameworks (Figure \ref{modular}). As we observe, most of the cases first two steps-- $S_{1}$ and $S_{2}$ can be executed with one component in  $\sum _{DP}$. These two steps can be combined into one $M_{DP}$. Other two steps-- $S_{3}$ and $S_{4}$ require more than one components in $\sum _{DP}$. Another key thing to remember is that in data-parallel components, input entities and parameters are a different thing (Smith and Albarghouthi \cite{mapreduce} define them as \emph{arity} and \emph{free variable} respectively). Sometimes, step $S_{4}$ requires the input parameters which value is calculated from either $S_{1}$ or $S_{2}$. Consequently, $S_{3}$ should be in a separate $M_{DP}$. Similarly, $S_{4}$ requires the input parameter calculated from all collective elements from $S_{3}$. Therefore, $S_{4}$ is separate from $S_{3}$. For many image processing tasks, $S_{3}$ and $S_{4}$ combined into a single data-parallel step. However, for reusable and customization perspective, we propose to wrap up the independent meaningful four steps into four $M_{DP}$-- $M_{DP}(S_{1}), M_{DP}(S_{2}), M_{DP}(S_{3}), M_{DP}(S_{4})$.
As we noticed in data-processing patterns, in many cases outcomes of the steps are required to flow and retain among intermediate steps (even up to the last step). That poses a challenge to data-parallel implementation as this data flow may increase both run-time memory and execution overhead. We present a solution considering a common list of data-entities with defined order to link-up data-flows among the $M_{DP}$. We recommend a three-layers vertical implementation of $M_{DP}$ for image pipelines which are presented in Figure \ref{mdp}. 
Layer-1 consists of abstract interfaces and $\sum _{DP}$, layer-2 handles parallel data-flow (DPF) and order of data entities (pseudo code is shown in Listing \ref{api_code}), and layer-3 contains $S_{i}$ on images. Data-parallel operations could be optimized using layer-1 without considering others. Layer-3  also works as a bridge to include image processing libraries (Skimage, OpenCV). Processing logic in this layer can be improved without the knowledge of layer-1. Components of Layer-1 call components in Layer-2, and Layer-2 call components in the lower layer. Therefore, three layers version of data-parallel module, $M_{DP}(S_{i}) = \sum _{DP} .>DPF.>\sum _{OS}[N,R]\{NP_{S}\}$. Here \emph{N} is the input entities (similar to RDD elements in Spark) populated by PABS, and \emph{R} is the list of parameters as described in data-processing patterns, $I$ and $I_{P}$ can be common in \emph{N}. Only $\sum _{DP}$ (via PABS) will call $\sum _{OS}$ through $DPF$. This modularity model provides a multidimensional (3x4, three layers and four modules) separation of concerns and dependency inversion principle (which is valuable for parallel development as distributed programming experts and image processing experts are not the same people usually). This will give the tool developer a common programming model to rapidly implementing the sequential tasks into Big Data platform. Finally, we recommend to save $I_{R}$ and $L_{M}$ into distributed storage, other types of result should be stored either in flat storage or databases. A block diagram is shown in Figure \ref{interface} on how common programming interface could be utilized using the $M_{DP}$ and processing patterns for interactive workflow/pipeline development. However, in the experimentation phase, we will discuss on what will be the impact of modularization and maintain a common list of data entities for each $M_{DP}$.

\section{Impact Analysis of Modularization}

In this section, we will discuss our experiments to observe the modularising effect of the image processing applications based on the modularity model presented in the previous section. We have three different Big Data infrastructure: (i) built with stand alone frameworks (Hadoop and Spark), (ii) built with Cloudera platform, and (iii) another is built with OpenStack. Big Data infrastructure built with Cloudera provides much flexibility of cluster setup and configuration, but our experimentation reveal it is a bit slower than stand-alone system. On the other hand, OpenStack facilitates dynamic node creation without further installing the frameworks and libraries (Cassandra, OpenCV, and so on) for each machine. In this paper, we present our experimental result with the second and third infrastructure cluster. The first infrastructure, Cloudera cluster consists of seven worker nodes (physical machines) with total 58 cores and 56GB RAM. The model of the processors is Intel Xeon L5420 and the speed is 2.50GHz. The master node is configured to 36GB main memory for the cluster driver. In ComputeCanada resource, our infrastructure is built by OpenStack with 5 instances each has 8 cores and 30GB RAM (Intel Xeon CPU E5-2650, 2.60GHz). This infrastructure is slightly better in terms of processor speed and main memory. Therefore, experiments with different configurations provide us the intuitive understanding of how the DCM behaves in different contexts. We conduct our experiment on four datasets: a set of CANOLA field images (most of them contain flowers, each image size is 1280x720), two sets of crop field images (each image size is 1280x960), and LSVRC2015 \footnote{http://image-net.org/challenges/LSVRC/2015/}. We implemented the programs with Spark-2.0.1 and Python 2.7. We compared the execution time between modularised and compact versions of six image processing tasks: (i) Image matching \cite{imgmatching}, (ii) Image classification \cite{imgclass}, (iii) CANOLA flower count (modified version of the base algorithm \cite{flowercount} for B-Channel), (iv) Object extraction \cite{segment_water}, (v) Image registration \cite{img_regi}, and (vi) Mosaic image generation \cite{panoramic}. We utilize the OpenCV \cite{opencv} image processing library to implement the operations in Layer-3 for the DCM. We found interesting behavior of differences between modularised and non-modularised versions. In the case of time-intensive operations, there is a significant performance issue when modularization is used for some tasks (i.e., Image Registration). However, in many cases, there is no significant differences such as object extraction, matching, and mosaic image generation.

From the experimentation with the Compute Canada infrastructure, shown in Table \ref{performance_computecanada}, we notice that the difference in execution time $\Delta(t)$ for counting flowers from 2K images, $T_{min}-T_{op}$ = 2.0 $-$1.9 = +0.1 minutes (+0.6 for slower cluster Table\ref{performance_little}), while for 8.6K (3.9GB) images, the modular version overcomes the run-time memory exceeding issue of the master node (with 30GB RAM). That means modularising the tasks facilitates performance optimization for individual step (we improved the performance by splitting runtime data at the third step according to the number of map partitions). For image registration, $\Delta(t)$ is about $-4$ minutes slower for both data size for the modular version, that means arbitrary modularization affects performance significantly. The performance is affected because of extensive data flow (image bundles) from the first data-parallel module to the last one. Whereas, the minimal step version has virtually no parallel data-flow. Likewise, for the larger dataset,  $\Delta(t)$ for modular image classification is $-7$ minutes (for first infrastructure in Table \ref{performance_little} this difference is  $-13$). All other cases, $\Delta(t)$ is almost 0, meaning no impact on execution time between modular and non-modular versions. We also experimented with the LSVRC2015 dataset for matching, clustering, and object extraction (as other pipelines require crops field images) and found that modularity does not impact execution time (presented in Table \ref{iccv_performance}) negatively for the large collection of images. On the other hand, for lower configured infrastructure, the time difference is higher despite enough resources.

However, we found some challenging tasks during our experiment. 
For the 2K images, the performance of the flower counting task is feasible, but we found that for the 8.6K dataset either the execution time is unusual or causing memory exceeding issues during runtime. Modularization and further splitting up the third data-parallel module of the flower counting task solved the memory issue, and the execution time decreases. However, in the mosaic image generation task, processing 300 images takes more than 400 minutes (while the well-configured machine takes 210 minutes for the usual program) for both the minimal and modular version; with the increment of images, the complexity arises (603 images take more than 600 minutes). Above all, although for a few tasks, modularity increases execution time, we can say that data-centric modularity increases the opportunity to further optimization for many cases.

\begin{table*}
    \small
	\centering
	\captionof{table}{Performance comparison (in minutes) of two versions of the image processing tasks in OpenStack infrastructure (ComputeCanada) (execution time is presented including I/O operations). Here min steps means minimal number of separate operations with map-reduce components.} \label{performance_computecanada}
	\small
	\begin{tabular}{p{.60 in}|p{0.35in}  |p{.50in}|p{0.50in} || p{0.35in}  |p{.50in}|p{0.50in} |p{0.20in}} \hline
		\textbf{Tasks} & \textbf{\#Img} & \textbf{$T_{min}$} & \textbf{$T_{mod}$}& \textbf{\#Img} & \textbf{$T_{min}$} & \textbf{$T_{mod}$}& \textbf{$N_{S}$ }\\\hline 
		IMatch	& 2K & 2.2& 2.2& 8.6K &9& 9 & 1 \\\hline
		Clustering	& 2K &6.6& 6.4& 4K & 15& 18 & 2 \\\hline
		FCount& 2K & \textbf{2}& \textbf{1.9} & 8.6K & \textbf{MemI}& \textbf{7.2}& 3\\\hline
		OBE	& 2K	& 0.30& 0.30& 8.6K & 0.83& 0.83 & 1\\\hline
		
		IMReg& 0.5K  & \textbf{8} & \textbf{12}& 1.5K & \textbf{25} & \textbf{32} & 1 \\\hline
		Mosaic	& 0.2K  & $>$300& $>$300& 0.3K & $>$400& $>$400 & 2\\\hline
		\multicolumn{8}{c}{ Here, $T_{min}$ -time for non-modular, $T_{mod}$ -time for modular, $N_{S}$ -steps for $T_{min}$} \\
		\multicolumn{8}{c}{FCount-Flower count, IMatch-Image Matching, OBE-Object Extraction} \\
		\multicolumn{8}{c}{IMReg-Image Registration, MemI-Memory issue} \\\hline
	\end{tabular}
	\vspace{-0.30cm}
\end{table*}

\begin{table*}
    \small
	\centering
	\captionof{table}{Performance comparison of two versions of the image processing tasks with Cloudera cluster.} \label{performance_little}
	\begin{tabular}{p{.60 in}|p{0.35in}  |p{.50in}|p{0.50in} || p{0.35in}  |p{.50in}|p{0.50in} |p{0.20in}} \hline
		\textbf{Tasks} & \textbf{\#Img} & \textbf{$T_{min}$} & \textbf{$T_{mod}$}& \textbf{\#Img} & \textbf{$T_{min}$} & \textbf{$T_{mod}$}& \textbf{$N_{S}$ }\\\hline 
		IMatch& 2K & 3.3& 3.3& 8.6K &13& 13 & 1 \\\hline
		Clustering	& 2K &11& 11& 4K & 18& 27 & 2 \\\hline
		FCount& 2K & \textbf{6.1}& \textbf{5.5} & 8.6K & \textbf{MemI}& \textbf{19} & 3\\\hline
		OBE	& 2K	& 0.8& 0.8& 8.6K & 2.1& 2.1 & 1\\\hline
		
		IMReg	& 0.5K  & \textbf{9.3} & \textbf{13}& 1.5K & \textbf{27} & \textbf{40} & 1 \\\hline
		Mosaic& 0.2K  & $>$300& $>$300& 0.3K & $>$400& $>$400 & 2\\\hline
	\end{tabular}
	\vspace{-0.30cm}
\end{table*}

\begin{table*}
    \small
	\centering
	\captionof{table}{Performance comparison of two versions of the image processing tasks with ILSVRC2015 dataset in OpenStack infrastructure (ComputeCanada) (total 40 cores and 150GB main memory )} \label{iccv_computecanada}
	\begin{tabular}{p{.60 in}|p{0.35in}  |p{.50in}|p{0.50in} || p{0.35in}  |p{.50in}|p{0.50in} |p{0.20in}} \hline
		\textbf{Tasks} & \textbf{\#Img} & \textbf{$T_{min}$} & \textbf{$T_{mod}$}& \textbf{\#Img} & \textbf{$T_{min}$} & \textbf{$T_{mod}$}& \textbf{$N_{S}$ }\\\hline 
		IMatch& 30K & 4.6& 4.6& 60K &8& 8 & 1 \\\hline

		OBE	& 30K	& 1.9&  1.9& 60K & 3.2& 3.2 & 1\\\hline

	\end{tabular}
	\vspace{-0.30cm}
\end{table*}

\begin{table*}
    \small
	\centering
	\captionof{table}{Performance comparison of two versions of the image processing tasks with ILSVRC2015 dataset in Cloudera infrastructure} \label{iccv_performance}
	\begin{tabular}{p{.60 in}|p{0.35in}  |p{.50in}|p{0.50in} || p{0.35in}  |p{.50in}|p{0.50in} |p{0.20in}} \hline
		\textbf{Tasks} & \textbf{\#Img} & \textbf{$T_{min}$} & \textbf{$T_{mod}$}& \textbf{\#Img} & \textbf{$T_{min}$} & \textbf{$T_{mod}$}& \textbf{$N_{S}$ }\\\hline 
		IMatch& 30K & 11& 11& 60K &22& 22 & 1 \\\hline

		OBE & 30K	& 3.3&  3.2& 60K & 6.5 & 6.5& 1\\\hline

	\end{tabular}
	\vspace{-0.30cm}
\end{table*}

Apart from these, if we follow common data processing patterns as described in Section \ref{data_pattern}, it is possible to write common data-parallel modules at the micro-level for the high-level components of large data analytic tools. This not only facilities re-usable module development but also we can write a common abstract interface to work with image processing without much knowledge of data-parallel operations and tuning. If separate operations are implemented as common method signatures within a class and its object instances are passed through corresponding data-parallel module, then processing logic can be reused or customized willingly without the knowledge of data-parallel components. With our model, the image processing tasks which contains only parallelisable operations in a processing step $S_{i}$ can be easily transformed into data-parallel programs without knowing the details of Layer-1 and Layer-2. Furthermore, if the input parameters for the DCM modules do not depend on the outcome of any of the steps in $S_{i}$ (and no non-parallel operations) then those DCMs for image processing tasks can be automatically converted into a data-parallel module. Wrapping all of these concepts, we also developed a library called SHIPPI which is being used by other programmers for developing and deploying image processing pipelines. Coupled with the data-processing pattern (in Section \ref{data_pattern}), our model represents a strong design rule \cite{leveragedr} in this domain. For instance, consider a project where one team is working on the web part, one team is working on the efficient large-scale data processing support, and another team is working on the image processing part; here dependency inversion principle (depend upon abstractions; do not depend upon concretions \cite{dependency_inversion}) is essential. We are working for more sophisticated techniques and algorithms to design a module to auto-transform the image pipelines into data-parallel modules, but this is out of the scope of our study. One critical observation is that for a map-reduce framework, developers should avoid arbitrary modularisation, unlike usual programming. Along with the large-scale data-handling and modules re-usability, our data-centric model explicitly supports the scenarios S3, S5, S6, S11, S12 described in \cite{cloud_archi}. We believe, this modularity model decreases the architectural changes of the core modules in a greater extent since a DCM senses and pre-defines most of the possibilities of a sub-domain as we have shown in the case study. 

\subsection{Challenges of Unified Data Interface for Large Scale Images}
IO and storage model is crucial for the performance of massive data processing. In order to solve the problems, research on unified data interface has been gaining traction \cite{diana}. However, we have not found extensive experimentation on image data for unified data interface. Here, we provide (only for large data among our many other IOs) our experience with various IO and storage model with the state-of-the-art technologies for our designed unified data interface for the data-centric modules. Primary options for the IO and storage model for large-scale images are: (i) Distributed storage (HDFS), (ii) Flat storage (both local and remote machine), and (iii) Database. IO with the distributed storage HDFS is faster but one problem is that processed images are saved as text data and another local program needs to convert it to images for making the result usable by the researchers. However, we experimented four types of model; one of the models we implemented is parallel loading and saving images via worker nodes utilizing SSH protocol which can handle up to 2K images only. The experimental outcome of other three models is presented in Table \ref{io_performance}. We found that IO for images are still challenging for feasible image analysis tasks, for instance, reading HDFS and writing to flat storage (master node) takes $\sim16$ minutes for 60K (5.7GB) images while HDFS-HDFS takes 84 seconds. However, despite more IO time, the benefit of HDFS-flat model is that the analysts get the ready-to-use processed images. On the other hand, for data and result management, Cassandra reveals the most effective data storage model (among other NoSQL storage, Cassandra is one of the most efficient models \cite{casperform, hadoop_cas} for Big Data). In summary, an effective I/O model is essential for the unified interface for handling large-scale images. Our data-centric module supports configurable storage-models at run-time for the storage of intermediate results as shown in Table \ref{modular}.
\begin{table*}
    \footnotesize
	\centering
	\captionof{table}{IO time for various data-sets. Here, Crp is crops field images; Flwr is crops flowers images (Cloudera cluster), Cndra - Cassandra. Both, HDFS and Cassandra replication factor are 1.} \label{io_performance}
	\begin{tabular}{p{.68 in}|p{0.60in}  |p{.60in}|p{0.60in} | p{0.60in}  |p{.60in}} \hline
		\textbf{IO Model} & \textbf{1.5k(Crp)} & \textbf{4k(Flwr)} & \textbf{8.6K(Flwr)}& \textbf{30K(VRC)} & \textbf{60K(VRC)} \\\hline \hline
		Cndra-Cndra& 12s &16s& 26s& 24s & 30s \\\hline
		Hdfs-Hdfs& 11s &16s& 25s& 46s & 84s \\\hline
		Hdfs-flat	& 1.4mins	& 4.8mins &  9.2mins & 8.1mins & 16.1 mins\\\hline

	\end{tabular}
	\vspace{-0.30cm}
\end{table*}

\bigskip

\begin{figure}

	\centering
	\vspace{-.3cm}
	\includegraphics[height=1.25in, width=3.6in]{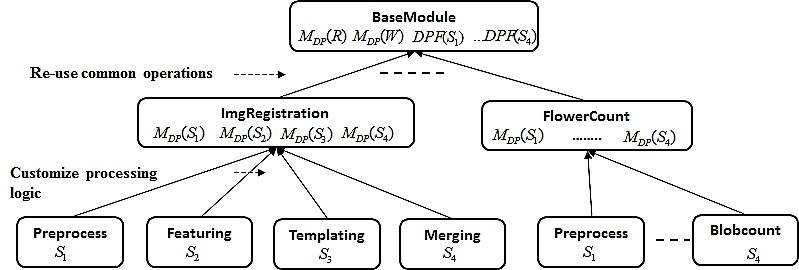}
	\small
	\caption{Options of reusability and customization. Common modules such as DPF can be placed in BaseModule and tasks (e.g., FlowerCount) can reuse those along with the other modules. Each canonical step ($S_{i}$) can be a separate module; they are reused and customized for each of the tasks with new computational logic without the knowledge of upper layer of ($M_{DP}$).}
	\vspace{-.4cm}
	\label{reusable}
	
\end{figure}

Apart from these, if we follow common data processing patterns, it is possible to write common data-parallel modules ($M_{DP}$) at the micro-level for the high-level components of large data analytic tools. This not only facilities re-usable module development but also we can write a common abstract interface to work with image processing without much knowledge of data-parallel operations ($\sum _{DP}$) and tuning. If separate operations are implemented as common method signatures (as shown in Figure \ref{interface} and Listing \ref{api_code}) within a class and its object instances are passed through corresponding $M_{DP}$, then processing logic can be reused or customized (shown in Figure \ref{reusable}) willingly without the knowledge of data-parallel components. With our model, the image processing tasks which contains only $\sum_{OS}$ in $S_{i}$ can be easily transformed into data-parallel programs without knowing the details of Layer-1 and Layer-2. Furthermore, if $R_{S1}$ to $R_{S4}$ do not depend on the outcome of any of the steps in $S_{i}$ (and no $NP_{S}$) then those image processing tasks can be automatically converted into  $M_{DP}$. Coupled with the data-processing pattern, it is logical to treat the model as a strong design pattern and design rule \cite{leveragedr} in this domain (for instance, consider a project where one team is working for the web part, one team is working for the efficient large scale data processing support, and another team is working for the image processing part; here dependency inversion principle is essential). More sophisticated techniques and algorithms might provide a framework to auto-transform the image pipelines into $M_{DP}$ in future which are out of the scope of our study. 

\begin{lstlisting}[language=Python,  label={api_code},caption=Pseudo code of DPF of $M_{DP(S_{2})}$ and $M_{DP(S_{3})}$]
DPFEstimate(N, obj, Rs2)
    unpack(N)-->im_id, I, Ip
    metrics = obj.estimate(Ip, Rs2)
    ...
    return pack(im_id, I, Ip, metrics)
DPFModel(N, obj, Rs3)
    unpack(N)-->im_id,I, Ip, metrics
    Dm = obj.model(Ip, metrics, Rs3)
    ...
    return pack(im_id, I, Ip, Dm)    

\end{lstlisting}

\section{Discussion}

\begin{lstlisting}[language=Python,  label={goodmosaic},caption=Splitted algorithm of Mosaic image with Spark]
Pr_RDD <-- RDDraw.map(preprocess)
Fs_RDD <-- Pr_RDD.map(faetureExtract)
Mosaic <-- Fs_RDD.first()
Tmp_RDD <-- Fs_RDD.zipWithIndex().cache()
#split = #elements / split_size
traversed.add(Mosaic)
do until all_images traversed
for i=0 to #split
    start = i * split_size
    end = start + split_size
    Filter_RDD <-- Tmp_RDD.filter(not(traversed) and in(start, end))    
    MFeature <-- broadcast(faetureExtract(Mosaic))
    Matched_RDD <-- Filter_RDD.map(matchpoints, MFeature)
    MaxImg <-- Matched_RDD.reduce(max(matchratio))
    if(MaxImg.ratio > previous_ratio)
        SeletedImg <-- MaxImg
        traversed.add(SeletedImg) 
    Mosaic <-- mergeHomography(Mosaic,SeletedImg)
result <-- Mosaic
\end{lstlisting}

\textbf{Lesson learned}: In summary, from our case study we extracted the following important insights:
\begin{itemize}
    \item Still, I/O operations create a bottleneck for optimal image processing with data-parallel frameworks 
	\item Considering modular data-processing patterns will reduce the implementation efforts and increase the reusability of both the program and the processed entities in data-parallel frameworks.
	\item Programmer should avoid arbitrary modularisation.
	\item Programmer should not rely on usual map-reduce concepts and tune hardware resources only for computation intensive tasks. 
	\item Intelligently splitting up the map-reduce operations and run-time data further might solve the limited resource problems as well as increase performance. 
	\item All image processing tasks may not be a good fit for traditional map-reduce techniques.
\end{itemize}

\section{Related Work}
A number of studies \cite{knnis, bigkmean, hadoop_mosaic, biggraph} have pointed out the challenges and problems of implementing the computation intensive tasks for scientific data with the abstract data-parallel frameworks in spite of having enough computing resources. To reduce the efforts of the data-scientists for large scale data analysis, some applications and frameworks are being developed \cite{spatialhadoop, hadoopbam, sparkseq, module_soft} for GeoSpatial and Bioinformatic data processing by adding more abstract layers on top of map-reduce frameworks. Despite enough progress, they do not support image processing operations. However, a few studies attempted to develop software and tools \cite{icp, iabdt, tomogram, imageharvest, dedip, kira, hipi} for large scale image processing for few specific cases. Nonetheless, they do not provide a common framework for diverse image processing pipelines. Our objective is to develop a scalable, unified and abstract framework for developing interactive image processing pipelines for large scale data.

Nowadays, large-scale images are used for analysis in various scientific works and general computer vision. Although a few studies provide techniques \cite{hipi, kira,stormcv, hadoop_mosaic, highregistration} for specific image processing tasks with data-parallel frameworks, they do not describe the challenges and optimization techniques to overcome the challenges. In this study, we highlight the challenges of real world large scale image processing tasks as well as recommend optimization technique with data-parallel components.

It is proven that program modularization is a key concept for developing unified frameworks on top of distributed and map-reduce programming environment. However, modularising map-reduce job (computation and data-intensive) is still challenging as data-parallel frameworks only provide a limit of few strict API methods. Yang et al. \cite{mrrunner} attempted to develop a framework for running modular map-reduce jobs, but users need to provide modularized jobs and dependency information. Recently, Heit et al. \cite{model_deployment} proposed a modular architecture for working with statistical models for data mining. However, these studies do not provide any technique of micro-level modularity and impact of modularizing the tasks. In our work, we propose a strategy for modularising large-scale image processing tasks at the micro-level and illustrate the pros and cons of modularising tasks with data-parallel frameworks.

In summary, our study on computation-intensive task analysis strategy, modularity model, and experimental insights will provide the researchers to focus on such challenges extensively for devising better techniques, and developers to consider the insights during large scale image processing tools development.

\section{Conclusion}
In this paper, we presented a case study on modularising data and computation intensive tasks into micro-level components. Our focus is on large image data as there is a lack of studies on the implications of running a wide variety of image processing tasks on Big Data platforms. We synthesize image data-processing patterns and propose a unified modular model for the effective implementation of computation-intensive tasks on data-parallel frameworks considering reproducibility, reusability, and customization. Our experimental results with six real world image processing tasks show that splitting and modularising the computation tasks is crucial to utilize the power of Big Data platforms. However, not all tasks show similar performance in execution time after modularising. A few of them need more sophisticated techniques for optimization with data-parallel frameworks. Therefore, our study provides a valuable knowledge-base for abstract and unified frameworks development for large scale data analysis. In future, we will work on techniques for automatic transformation of sequential tasks to data-parallel modules.





\footnotesize
\bibliographystyle{IEEEtran}

\bibliography{IEEEbig}
%

\end{document}